\documentclass[prl,floatfix,twocolumn,showpacs]{revtex4}
\usepackage{amsmath}
\usepackage{amsfonts}
\usepackage{mathrsfs}
\usepackage{epsfig}
\usepackage{graphicx}
\usepackage{dcolumn}

\begin{document}
\title{Modularity promotes epidemic recurrence}
\author{Tharmaraj Jesan$^{1,2}$, Chandrashekar Kuyyamudi$^2$ and Sitabhra Sinha$^2$}
\affiliation{$^1$ Health Physics Division, Bhabha Atomic Research
Center, Kalpakkam 603102, India.\\
$^2$ The Institute of Mathematical Sciences, CIT Campus,
Taramani,
Chennai 600113, India.}

\date{\today}
\begin{abstract}
The long-term evolution of epidemic processes depends
crucially on the structure of contact networks.
As empirical evidence
indicates that human populations exhibit strong community
organization, we investigate here how such mesoscopic
configurations affect
the likelihood of epidemic recurrence.
Through numerical simulations on real social networks and
theoretical arguments using spectral methods,
we demonstrate that highly contagious diseases that would have
otherwise died out rapidly can persist indefinitely for an optimal
range of modularity in contact networks.
\end{abstract}

\maketitle

\newpage

Infectious diseases continue to remain among the major causes of
human mortality worldwide, despite considerable progress in their
treatment and management~\cite{WHO2004}. In recent decades,
periodically reemerging epidemics have posed a significant
challenge to public health globally~\cite{Morens04,Fauci05}.
Various causative factors for such re-emergence have been suggested,
including
zoonotic encounters~\cite{Jones2008,Morse2012}, environmental
degradation~\cite{Dobson2001} and
periodic variations in climate~\cite{Kovats03,Sultan05}.
However, such explanations are crucially dependent on exogenous
factors specific to a particular outbreak.
A more general framework for explaining the recurrent pattern of 
epidemics should involve
endogenous properties of contagion spreading between
individuals in a population. Critical determinants
of such processes include the properties of the contact network
that allow an infection to
propagate~\cite{Kitsak2010,Goltsev2012,PastorSatorras2015}. 
More generally, 
explaining
how long-term recurrence can arise in dynamical systems coupled by 
non-local diffusive interactions can contribute to understanding
persistence in non-equilibrium systems~\cite{Schehr2007,Majumdar2010,Bray2013}.


A prominent topological characteristic of social networks is their modular
nature~\cite{Wasserman94,Girvan02,Onnela07,Nematzadeh2014}:
it is possible to 
identify {\em communities} with a high density of connections
between their members,
as compared to those between members of different
communities. 
Earlier studies have shown that diseases
are less likely to become established in
networks that are strongly
modular~\cite{Liu05,Huang06,Huang07,Zhao07,Griffin12}. 
It has also been
shown that immunization in modular networks is more effective if
individuals bridging communities, rather
than the most highly connected individuals, are preferentially
targeted~\cite{Masuda09,Salathe10}.
However, the impact of such mesoscopic organization of populations on
the eventual fate of a contagion has remained
relatively unexplored. This is surprising given that the long-term
outcome of an epidemic breakout is critical from the perspective of
disease eradication and control. 
In particular, the study of
persistence time of a disease~\cite{Earn98} 
reveals the existence of a critical population 
size~\cite{Bartlett57} 
below which an epidemic,
after an initial phase of rapid growth, becomes extinct due to the paucity
of susceptible individuals.
Theoretical
studies of disease persistence typically consider homogeneous 
random mixing in populations,
and it is an open question as to whether the presence of modular topological
organization in contact networks, as seen in human
societies, can significantly affect the critical population size 
for a highly infectious disease.

In this paper, we demonstrate the key role played by the
mesoscopic structure of the contact network, viz., its modular
organization, in the dynamics of epidemics at long time-scales. We
specifically
show that highly infectious diseases that would have otherwise died
out rapidly can persist indefinitely for an optimal range of
modularity. 
The critical role played by the mesoscopic structural organization in
this phenomenon is established by implementing stochastic
contagion spreading dynamics on an ensemble of empirical social
networks,
as well as on contact network models whose modularity can be tuned.
We observe that modularity differentially affects the fate of
infections depending on their contagiousness, quantified by the
basic reproduction number $R_0$, i.e., the average number of secondary
infections resulting from a single infected agent in the initial stage
of the epidemic when almost the entire population is 
susceptible~\cite{Anderson91}.
Thus, while epidemics with lower $R_0$ can persist in populations
exhibiting relatively homogeneous contact networks, those with 
higher $R_0$ can survive only
in networks with strong modular organization. This has
obvious public health policy implications, especially in designing
effective intervention strategies for countering recurrent epidemics.

The quantitative framework for understanding recurrence-driven
persistence of epidemics
is provided by the 
SIRS compartmental model of epidemic spreading~\cite{Anderson91}.
A population of $N$ agents (represented by the nodes of a network) 
is composed of 
susceptible ($S$), infected ($I$) and recovered ($R$)
individuals, whose
numbers vary with the progress of the
epidemic over time.
Each link between a pair of nodes is a
contact along which infection can propagate,
with the rate of infection transmission from an infected to a
susceptible agent being $\beta$.
Infected agents are assumed to recover at a rate $\gamma = 1 /
\tau_I$, where $\tau_I$ is the average
duration of the infection. 
A recovered agent has temporary immunity to the infection for a
period whose average value is $\tau_R$, after which it becomes susceptible 
again.
Thus, the
transition from recovered state to susceptible state occurs at
a rate $\mu = 1/ \tau_R$. Note that limiting cases of the SIRS
model yield other well-known models such as the SI
($\tau_I \rightarrow \infty$), SIR ($\tau_R \rightarrow
\infty$) and SIS ($\tau_R \rightarrow 0$), which have all been
used in studies of contagion propagation.
In a contact network with average degree (i.e., links per node) $k$, 
the basic reproduction
number for an SIRS process can be approximated as $R_0 \simeq k \beta
\tau_I$~\cite{Keeling00}.

To perform stochastic simulations of contagion propagation on a
contact network, we have used the Gillespie
direct algorithm~\cite{Gillespie77}. The rates of $I
\rightarrow R$ and $R \rightarrow S$ events are not constant, but
depend on the time elapsed after infection
($\Delta t_I$) and recovery ($\Delta t_R$) respectively,
viz., $\gamma = {\rm exp} (\eta (\Delta t_I - \tau_I))$ and $\mu =
{\rm exp} (\eta (\Delta t_R - \tau_R))$ where $\eta$ governs the
nature of the distributions of durations of infection and recovery of
individuals.
This is to take into account the fact that in reality the rate at
which an individual recovers in a given time interval
is initially small but increases over time, corresponding
to the infectious period distribution being less skewed than an
exponential distribution (which would have been obtained had the
recovery rate been assumed to be constant in time)~\cite{Keeling97,Lloyd01}.
We carry out simulations for durations (typically $>10^3$ time units) that
are much longer than any of the time-scales ($\tau_I$, $\tau_R$) governing the dynamics of
the SIRS model. While results shown here
are for $\tau_I =5, \tau_R = 10$, we have explicitly verified that
qualitatively similar behavior is seen for different choices for these
parameters.

\begin{figure}
\begin{center}
\includegraphics[width=0.99\linewidth,clip]{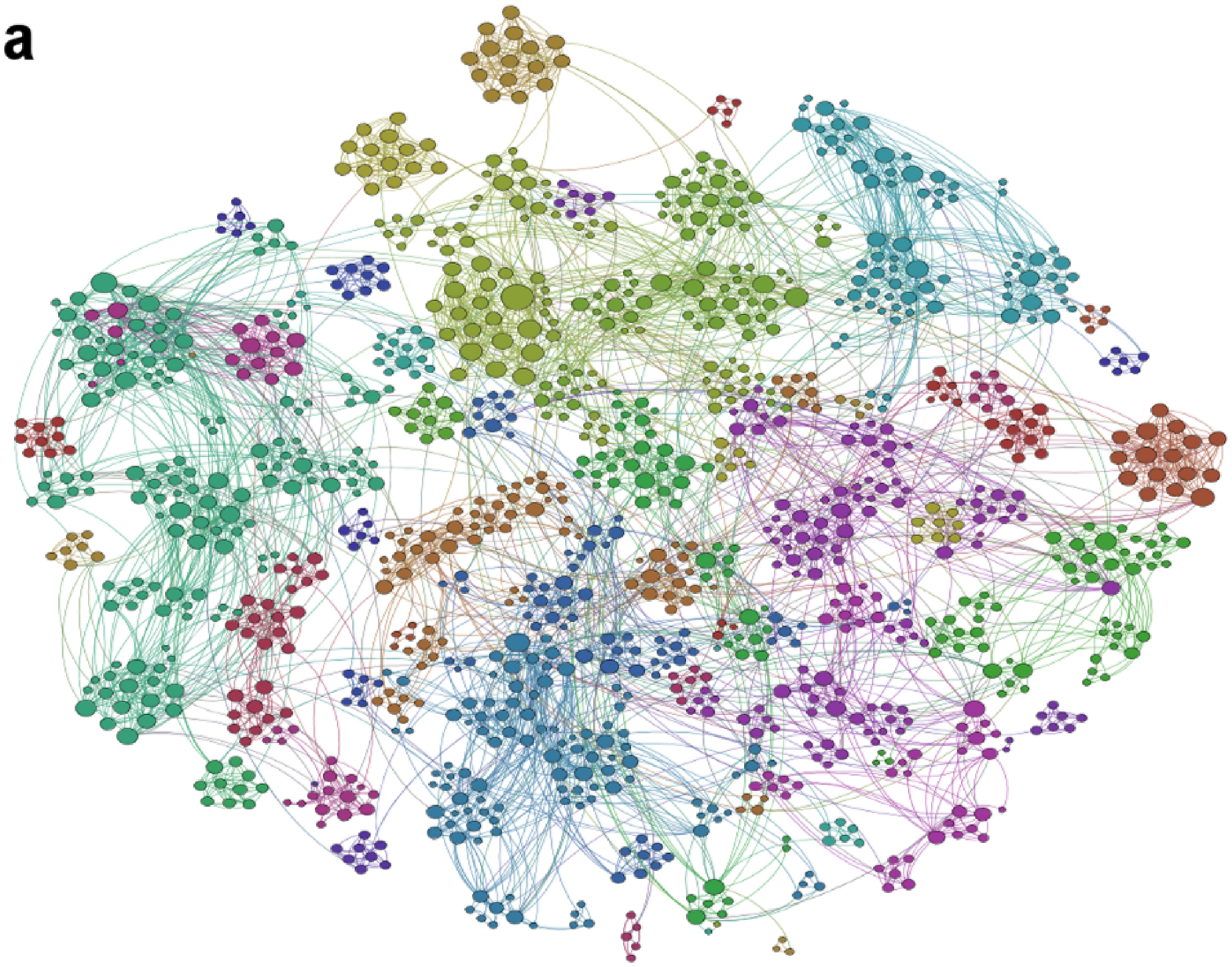}
\includegraphics[width=0.99\linewidth,clip]{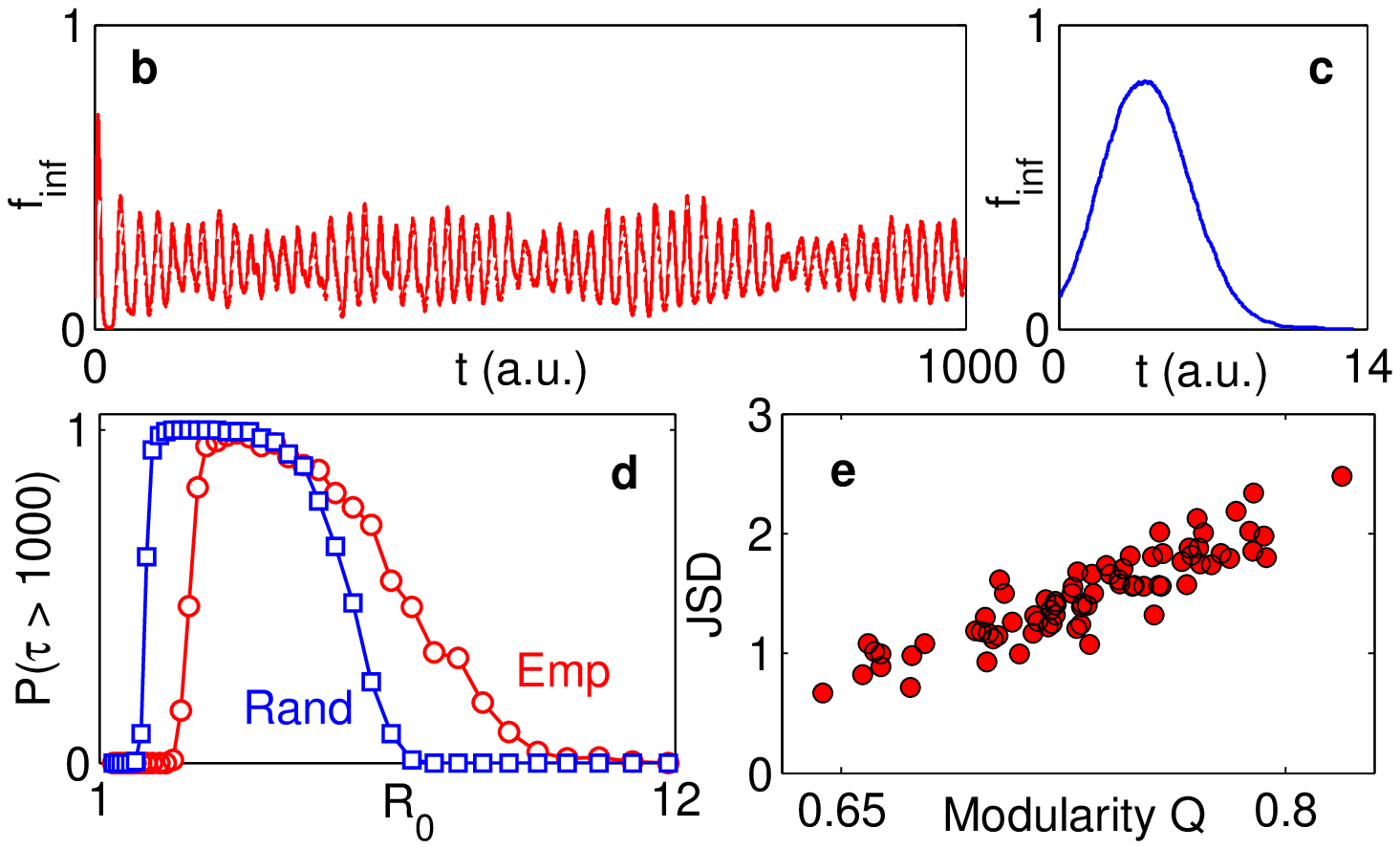}
\end{center}
\caption{(color online). 
(a) Visualization of the largest connected component of 
the social network in a village of
southern India comprising $N = 1151$ individuals (contact data
from Ref.~\cite{Banerjee13}), on which we simulate contagion
propagation.
Nodes are colored according to the
community they belong to, obtained
using a spin glass energy minimization
method~\cite{Reichardt06}.
(b-c) The time-evolution of the fraction of infected individuals,
$f_{inf}$, for a
simulated epidemic with basic reproduction number $R_0= 7$, started
by infecting a randomly chosen 1\% of the population in
(b) the empirical contact network shown above, and (c) a degree-preserved
randomization of the same network.
Note that the contagion is persistent for an extremely long period 
($>10^3$ time units) in the empirical
network compared to the randomized network where
community structure is absent.
(d) The probability that an epidemic persists for longer than
a specific duration 
in the empirical
contact network (red circles) and corresponding
degree-preserved randomized networks (blue squares, averaged over 100
network realizations) shown as a function
of the basic reproductive number $R_0$ of the epidemic. When the
contagion is more infectious ($R_0>5$),
the epidemic is more likely to be persistent in the empirical network 
compared to the randomized ones.
(e) The role of modularity (measured by $Q$) in increasing the
probability of persistence of an epidemic, measured as the 
Jensen-Shannon divergence between the persistence probability
distributions for empirical contact networks for 75 different villages
and that of corresponding degree-preserved randomized networks. The
linear correlation coefficient for correlation between $Q$ and $JSD$
is $0.89$ ($p$-value $= 0$).
}
\label{fig1}
\end{figure}

We first establish the crucial role played by modularity
in the long-term behavior of an
epidemic by simulating contagion spreading
on a set of empirical social networks. We have reconstructed these networks from
detailed information on social
interactions between individuals belonging to 75 villages in southern
India (using data from Ref.~\cite{Banerjee13}). 
Fig.~\ref{fig1}~(a) shows a representative network
corresponding to one of the
larger villages where the different modules
identified by a community detection algorithm~\cite{Reichardt06}
are indicated.
We observe that initiating an epidemic in the network results in
the contagion surviving indefinitely over a range of simulation
parameters [Fig.~\ref{fig1}~(b)]. However, if the network is
randomized so as to remove the modular structure
the contagion is extinguished rapidly in the same
parameter regime [Fig.~\ref{fig1}~(c)].
Fig.~\ref{fig1}~(d) shows how the persistence behavior of the
epidemic, measured in terms of the duration $\tau$ for which the
infection survives in the
population, differs between the empirical network and the corresponding 
randomized network ensemble -
the former being more likely to exhibit persistence of {\em highly
infectious} (viz., $R_0 > 5$) contagia than the latter.
As the randomized networks have a degree sequence identical to the
empirical network, it
suggests that the enhanced persistence of the epidemic in the latter
is a consequence of its modular organization. 
This is further supported by observing that the persistence behavior
is enhanced if we reduce only the inter-modular
connections in an empirical network (which effectively increases its
modularity~\cite{supp}).
Indeed, a quantitative relation between 
persistence behavior and modularity can be obtained by showing how the
difference in the persistence probability distributions for empirical
and randomized networks vary as a function of the modularity ($Q$) of the
empirical networks~\cite{Newman2006}.
This difference is measured by the Jensen-Shannon
divergence~\cite{Lin1991}, defined
for a pair of discrete probability distributions $X$ and $Y$ 
as:
$$JSD (X,Y) = \frac{1}{2} \Sigma_i \left( X_i \ln \frac{X_i}{Z_i} + Y_i \ln
\frac{Y_i}{Z_i} \right),$$ 
where $Z = \frac{1}{2} (X+Y)$. 
Fig.~\ref{fig1}~(e) shows that JSD increases almost linearly
with $Q$ of the contact
network, establishing the
critical role of modularity in enhancing the persistence of highly
infectious epidemics.

\begin{figure}
\begin{center}
\includegraphics[width=0.99\linewidth,clip]{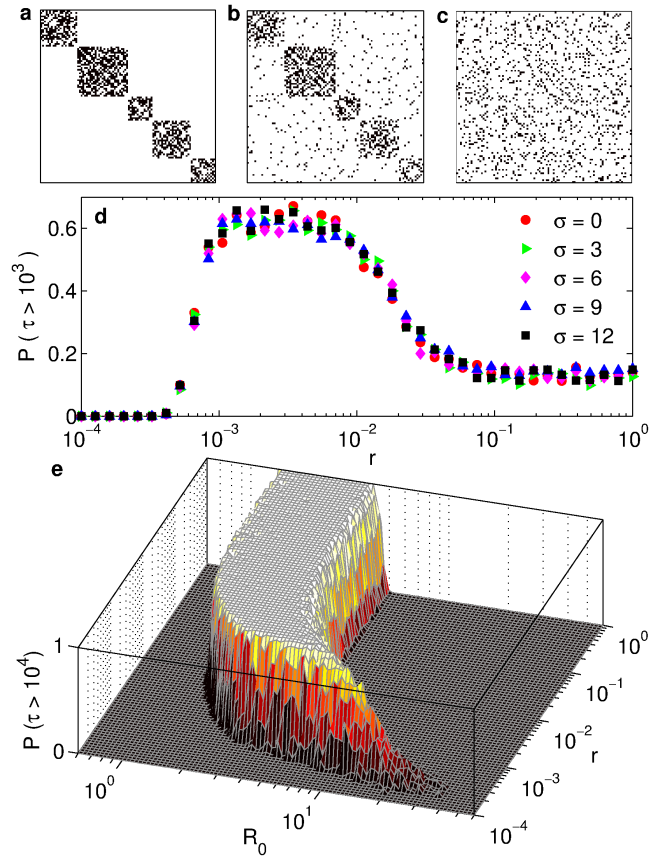}
\end{center}
\caption{(color online). 
(a-c) Adjacency matrices of the contact network model shown for
different values of the modularity parameter $r =
\rho_{out}/\rho_{in} \in [0,1]$, which is the ratio of inter-modular to
intra-modular connection density for a fixed average degree $k$.
Starting from a collection of isolated clusters
(a, $r=0$), by increasing $r$ we obtain modular networks (b, $r=0.1$)
eventually arriving at a homogeneous network (c, $r=1$).
(d)~The probability of persistence for an epidemic (viz., duration $\tau >
10^3$ time units)
having a basic reproduction number $R_0 = 6$ in a
modular network of $N = 1000$ agents ($k = 12$),
grouped into $M=20$ modules peaks in an optimal range of $r$,
independent of the heterogeneity in community sizes (measured by
their standard deviation $\sigma$). Increasing the number of modules
$M$ enhances the observed persistence behavior~\cite{supp}.
(e)~The probability of persistence ($\tau > 10^4$) of an epidemic in a
modular network ($N=1024$, $M=64$, $\sigma = 0$, $k=12$) as a function of
$R_0$
and $r$. In a homogeneous network ($r=1$), highly
infectious contagia (viz., $R_0 > 5$) rapidly infect almost all
agents and become extinct. Systems with almost isolated modules ($r
\rightarrow 0$) also exhibit short-lived epidemics as a local outbreak
in one module is unable to transfer to the others.
However, for an optimally modular
network ($r \sim 10^{-3}$), epidemics with high values of $R_0$
can become recurrent and
persist in populations for extremely long times.
}
\label{fig2}
\end{figure}
In order to investigate the mechanism by which community organization
affects the long-term dynamics of epidemics, we consider an ensemble of
model contact networks whose modularity can be tuned~\cite{Pan09}. 
Each network is
constructed such that the $N$ nodes (representing agents) comprising
the system are arranged into $M$ modules that can have varying
sizes. The size distribution has 
a Gaussian form with a mean size $\langle n \rangle (= N/M)$
and standard deviation, $\sigma$, which is a free parameter
that determines the heterogeneity in community sizes~\cite{Menon15}.
Unlike most earlier studies of epidemic dynamics on
community-structured networks, the modular nature of this
network model can be varied
continuously by tuning the ratio of inter- to intra-modular
connectivity, $r = \rho_{out} /\rho_{in} \in [0,1]$ without changing the
average
degree $k$ of the nodes [Fig.~\ref{fig2}~(a-c)].

Fig.~\ref{fig2}~(d) shows the probability that a highly
infectious contagion ($R_0 = 6$) survives for long times in contact
networks as a function of their modular character, parameterized by $r$.
We observe that
there is an optimal range of $r$ over which the 
epidemic becomes
persistent, independent of the level of heterogeneity in module sizes. 
This is consistent with the results obtained for empirical social
networks above, as the modular
nature of the contact
network is most prominent over this range while still keeping the entire network
connected.
Decreasing $r$ further makes the modules effectively isolated
from each other, preventing a local epidemic outbreak from spreading
to the rest of the population. On increasing $r$ the modular
structure become less prominent and the epidemic rapidly spreads
through the relatively homogeneous network. Both scenarios result in
the extinction of the infection in the population within a short
duration.

To see how this relation between persistence and modularity is affected
by the contagiousness of the epidemic - an important intrinsic
property of its dynamics -
in Fig.~\ref{fig2}~(e) we look at the
joint dependence of the probability that the
epidemic persists for long times on $R_0$ and $r$. The contagion 
results in an epidemic when $R_0 > 1$, and 
in the extreme limit where no modularity is apparent (i.e.,
the completely homogeneous network obtained for $r=1$) it will persist
in the population only if it propagates sufficiently slowly, so as to
allow recovered agents to become susceptible again while the infection
is still present in the system. Thus, we observe the probability of
survival of the contagion to be high only for epidemics with low
contagiousness ($R_0 <5$) at the limit $r=1$, which can be explained
quantitatively in terms of a solution of a delay difference
equation~\cite{supp}.
The range of $R_0$ over which the epidemic is seen to become
persistent remains effectively unchanged as $r$ is decreased
until the network becomes sufficiently modular [around $r \sim
10^{-2}$ in Fig.~\ref{fig2}~(e)], after which we observe a
gradual shift of the persistence range towards higher values of $R_0$.
However, if $r$ is decreased further, the modules eventually
become isolated [$r \gtrsim 10^{-4}$ in Fig.~\ref{fig2}~(e)]. This results in rapid extinction of any epidemic that is
initiated, as the contagion is unable to spread through the
entire population regardless of $R_0$.

%

%

To investigate in more detail the mechanism by which the lifetime
$\tau$ of an epidemic with high $R_0$ diverges in optimally modular networks, we
can examine how the nature of
its distribution $P(\tau)$ changes upon varying the
parameter $r$ controlling the modularity of the network.
Decreasing $r$ from $1$, as we approach the range where persistence is
observed the distribution becomes bimodal, splitting into two
branches~\cite{supp}. The lower branch corresponds to epidemic events 
confined within a community, where they spread rapidly and
become extinct when no further susceptible agents are available.
By contrast, the upper branch is peaked around a value
that diverges with decreasing $r$ and represents realizations where
the epidemic is able to successfully break out from the module in
which it started, subsequently spreading from one community to
another.

\begin{figure}[tbp]
\begin{center}
\includegraphics[width=0.99\linewidth,clip]{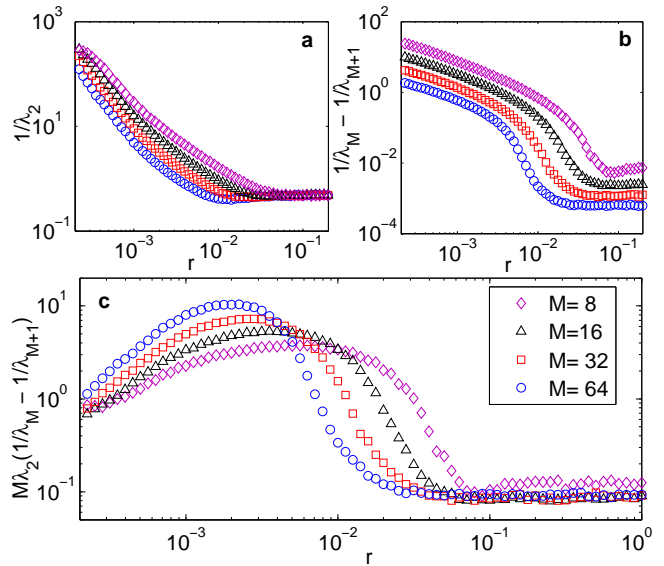}
\end{center}
\caption{
(color online). 
(a) Variation of the inverse of the smallest finite
eigenvalue of the Laplacian for a modular
network, representing the global diffusion time-scale, with
the parameter $r$. (b) The Laplacian spectral gap, corresponding to
the time-scale
separation between intra- and inter-modular processes, shown as a
function of $r$.
The different curves coincide on scaling with number of modules $M$.
The ratio of the time-scale separation (scaled by $M$) and global time-scale
is shown in (c), which exhibits a peak in the region corresponding to
optimal modularity for epidemic persistence.
The symbols represent networks
having different values of $M$ [see key in (c)].
For all results shown here $N=1024$ and $k=12$.
}
\label{fig4}
\end{figure}
Thus, the key for understanding epidemic
persistence in networks having community organization
is the occurrence of spreading involving different time-scales
in modular contact networks.
This can be understood analytically in terms of a diffusion process
defined on such networks with the probability of transmission from
agent $i$ to $j$ in a single time step being $p_{ij} = A_{ij}/k_i$,
where $A$ is the network adjacency matrix (i.e., $A_{ij} =1$, if $i,
j$ are connected and $=0$, otherwise) and $k_i$ is the
degree of $i$-th agent. The transmission probability matrix {\bf p} is
related to the normalized Laplacian of the network,
${\mathscr L}=$ {\bf I}$-${\bf D}$^{\frac{1}{2}}${\bf p}{\bf
D}$^{-\frac{1}{2}}$ where {\bf I} is the identity matrix and {\bf D}
is a diagonal matrix with $D_{ii} = k_i$. The 
Laplacian spectrum for undirected networks (as considered here)
comprises non-negative real eigenvalues
$\lambda_1=0 \leq \lambda_2 \leq \lambda_3 ....\lambda_N$.
The reciprocal of the eigenvalues are related to time-scales for
diffusion occurring over different ranges, with the smallest non-zero
eigenvalue ($\lambda_2$ for a
connected network) corresponding to
global diffusion, i.e., the contagion spreading through the entire
network.
Fig.~\ref{fig4}~(a) shows
that this time decreases with increasing $r$ as the network becomes
more homogeneous.
The role of the modular organization becomes clear if we focus on the
gap in the eigenvalue spectrum between $\lambda_M$ and $
\lambda_{M+1}$ for a network with $M$ modules, that is related to the
difference in the time-scales of fast
intra-modular processes and slow inter-modular processes~\cite{Pan09}.
Fig.~\ref{fig4}~(b) shows
that this spectral gap also decreases with increasing $r$, i.e., the
difference between the time-scales also increases.
The occurrence of persistence in an optimal range of
modularity is explained by looking at the ratio of the spectral gap
to the global diffusion time-scale
shown in Fig.~\ref{fig4}~(c). 
This quantity peaks for an
intermediate value of $r$ (between $10^{-3}$ and $10^{-2}$ in
Fig.~\ref{fig4}~(c)), 
where the epidemic propagates sufficiently rapidly
so as not to become extinct before being able to hop from one
module to another. Nonetheless, the global diffusion time is long
enough such that when the epidemic
returns after circulating around the network, a sufficient number of
agents would have become susceptible for the epidemic to continue. 


To conclude, we have shown that there is an optimal range of
modularity of the contact network for which highly contagious diseases
can persist indefinitely in the population.
Our work has potential ramifications for the long-term
management of contagious diseases, suggesting that determination of the critical
population size required for epidemics to become
recurrent has to necessarily take into account
the mesoscopic organization of the social network.
In particular, it suggests that interventions such as
quarantine that isolate local communities from each
other (thereby increasing modularity)
can have very different efficacies
from the perspective
of the long-term recurrence of a disease depending on its $R_0$.

We thank Shakti N. Menon, Raj K. Pan and Rajeev Singh for helpful
discussions, R. Janaki for assistance and IMSc for providing computational
resources.
This work was supported in part by IMSc Complex Systems Project
(XII Plan) funded by the Department of Atomic Energy,
Government of India.

\pagebreak

\begin{center}
{\large \bf SUPPLEMENTARY MATERIAL}
\end{center}
\vspace{0.25cm}

\setcounter{figure}{0}
\renewcommand\thefigure{S\arabic{figure}}  
\renewcommand\thetable{S\arabic{table}}  
\renewcommand\theequation{S\arabic{equation}}

{\em Data description.} Empirical social contact networks among
individuals residing in each of 75 villages located in southern
Karnataka, a state in the south of India, were reconstructed from
detailed information about the relationships of the villagers with
each other~\cite{data1}. 
This data was collected through a survey carried out as part
of a study on the operational feasibility of a
microfinance institution in these villages~\cite{Banerjee}. 
The different types of relations between the surveyed individuals that
were recorded included kinship, social engagement, mentoring, visiting homes,
borrowing and lending money or essential items, etc. 
We assume that the existence of any kind of
relation between a pair of individuals provides a pathway for the
transmission of the contagion. Furthermore, from the perspective of
spreading of a pathogen, the directional nature of a relation is
unlikely to be relevant. For our study, we therefore
consider the undirected network obtained from the union of all these 
relations between the individuals in a village as a representation of
the social contact network along
which an epidemic can spread in the village. 
Table~\ref{table1} gives the summary details of the contact network
properties for each of the 75 villages. It mentions the number of
individuals surveyed in each village (Nodes), the size of the
largest connected component, i.e., the largest group of nodes such
that between any pair a connected path can be found (LCC), the
average degree, i.e., the mean
number of connections for each node in the LCC ($\langle k \rangle$),
the number of communities (Modules) identified in the LCC using a spin glass
energy minimization method~\cite{Reichardt}, the
modularity measure $Q$ that quantifies the degree of community
organization in the LCC, the mean value ($\mu_{sz}$) and standard
deviation ($\sigma_{sz}$) of the sizes of modules in the LCC, the size
of the largest module in the LCC (LMS) and the ratio of the
inter-modular to intra-modular connection density for the LCC ($r$).

\begin{figure}
\begin{center}
\includegraphics[width=0.9\linewidth]{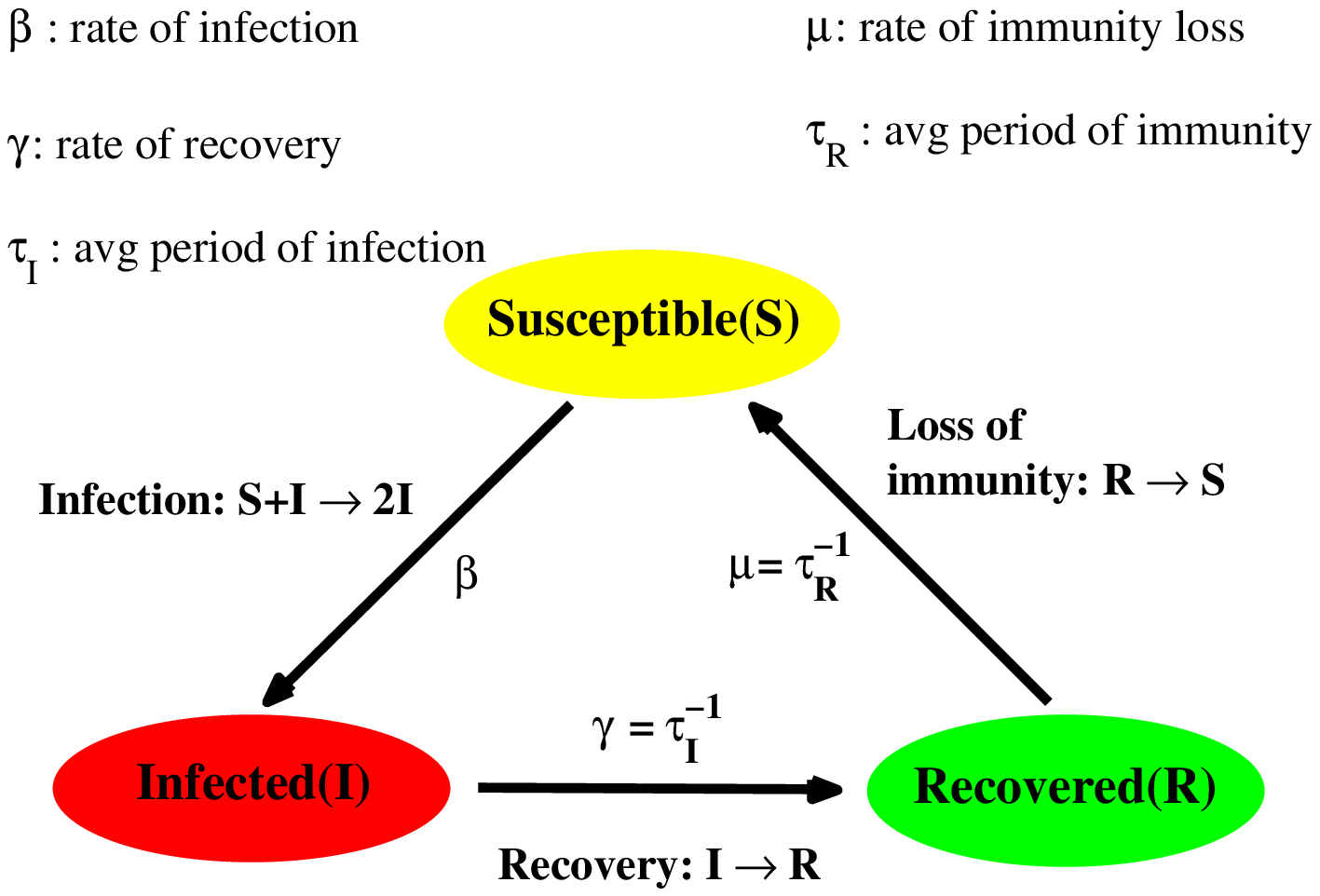}
\end{center}
\caption{Schematic  illustration of the dynamics of the SIRS model.
The  dynamical processes underlying the transition between the
different states, viz., Susceptible ($S$), Infected ($I$) and Recovered
($R$) are shown. Initially most of the population is in the compartment
$S$.
Passage of a few individuals to the compartment $I$ results in
more individuals transferring from $S$ to $I$ over time at a rate
$\beta$. At the same time, individuals move from compartment $I$ to
$R$ as they recover at the rate $\gamma$ (= 1/$\tau_I$, where $\tau_I$
is the mean period of
infection). Over long times, individuals move from $R$ back to $S$ because
of loss of immunity at the rate $\mu$ (= 1/$\tau_R$, where $\tau_R$ is
the mean recovery time).
}
\label{figs1}
\end{figure}

{\em SIRS Model.} A common mathematical framework for studying
dynamics of contagia spreading in a population is that of
compartmental models. Such models divide individuals belonging to a
population into several compartments corresponding to differing states
of health. For example, individuals could belong to the susceptible
($S$) to infection category, infected ($I$) with pathogen
category or in the category of recovered ($R$) from infection. In
addition, such a model also has to specify the probabilities (or
rates) at which individuals move from one category to another.
In our work to study the recurrence of epidemics we use the SIRS
model,
where individuals move from $S$ to $I$ following infection (with rate
$\beta$), from $I$ to $R$ following recovery (with rate $\gamma$) and
from $R$ to $S$ following loss of immunity (with rate $\mu$)
[Fig.~\ref{figs1}]. This allows contagia to re-enter the population
repeatedly.

Note that in the SIRS model, the immunity conferred on agents
immediately after recovering from an infection is only temporary.
Thus, after a certain
duration, individuals who have previously been infected re-enter the
susceptible compartment.
This may arise, for example, because (i) for certain diseases
such as pertussis and syphilis, affected individuals gradually
lose immunity and eventually again become susceptible to
infection~\cite{Konig95,Grassly05}, (ii) pathogens may undergo genetic
changes so that previously affected hosts immune to the original
strain are exposed to risk of infection from the novel
strain, e.g., as in influenza~\cite{Pease87,Hay2001,Hooten2010} and (iii) demographic
processes through which recovered individuals die and are replaced by
birth of susceptible individuals in the population (maintaining the total
population constant)~\cite{Hethcote76}.

{\em Details of stochastic simulation of epidemics on networks.}
We perform stochastic simulation of SIRS dynamics on the empirical and model
contact networks using the
Gillespie stochastic simulation algorithm~\cite{Gillespie}. 
There are three possible types of events that can occur in the network
at any time:
(i) A node (agent) that is in susceptible ($S$) state can undergo
transition to infected ($I$) state. The propensity of a particular node in $S$
state to move to $I$ state is
$\beta\,\frac{k_{inf}}{k}$, where $k_{inf}$ is the number of infected
neighbors of the node while $k$ is the total
number of its neighbors (i.e., degree).
(ii) A node that is in $I$ state can undergo
transition to recovered ($R$) state. Note that the process of an individual
node recovering from infection is independent of its
neighborhood.
(iii) A node that is in $R$ state can undergo
transition to $S$ state. The process of an individual
node in $R$ state moving back to $S$ state is also
independent of its neighborhood.  

To simulate the stochastic time-evolution of an epidemic process in a
network of $N$ nodes we need to determine (a) the
sequence of time instants at which a transition event occurs
and (b) the identity of the node at which the transition occurs. The
latter
determines the nature of the transition (viz., $S \rightarrow
I$, $I \rightarrow R$ and $R \rightarrow S$), as this depends on the
current
state of the node ($S$, $I$ or $R$).
The time-interval $\Delta{t}$ between two successive
transition events is calculated as $\Delta{t} =
-\log(\xi)/\Sigma^N_{i=1} p_i$, where
$p_i$ are the rates associated with each of the $N$ transition events
and $\xi \in [0,1]$ is a uniformly distributed random number in the
unit interval. After this interval is obtained, we determine which
one of the $N$ possible events will actually take place by generating
another uniformly distributed random number $\zeta \in
[0,\Sigma^N_{i=1} p_i]$ and pick the event $j$ that satisfies
$\sum_{i=1}^{i=j-1}{p_i}< \zeta \leq \sum_{i=1}^{i=j}{p_i}$ ($j = 1 ,
\ldots, N$). This process is repeated until either (a) there are no
infected individuals in the population such that the disease is
declared extinct, or (b) the maximum time allowed for the simulation is
reached.

\begin{figure}[htbp]
\includegraphics[width=0.99\linewidth,clip]{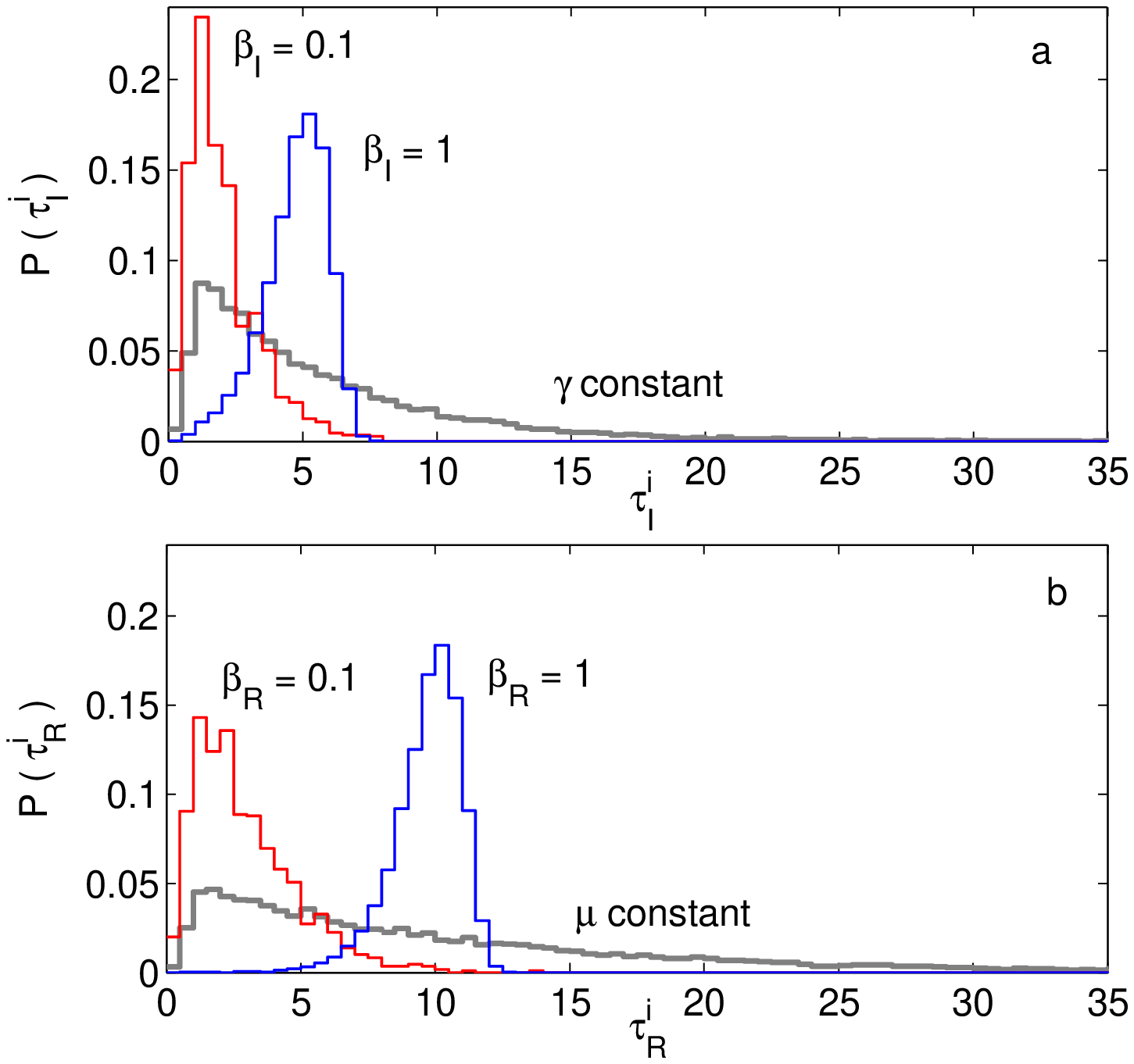}
\caption{The distribution of times for which an individual remains
infected (top) and the period for which an individual is in
the recovered state (bottom) in the stochastic model
simulations. The different curves correspond to simulations performed
with the rates of recovery ($\gamma$) and loss of immunity ($\mu$)
kept constant (red) or time-varying (green and blue curves
corresponding to model parameter $\beta = 0.1$ and $1$ respectively).
The distributions become more narrowly peaked about the respective
mean values [viz., $\tau_I$ = 5 (top), $\tau_R = 10$
(bottom)] with increasing values of $\beta$. 
Compared to the constant rate process (red curve), the
probabilities of having extremely short or long periods of infection and
recovery are markedly reduced for time-varying rates, especially as
$\beta$ is increased. Note that $\beta = 1$ for the stochastic
simulation results shown in this paper.
}
\label{timedistrn}
\end{figure}
Conventional stochastic simulations assume constant
rates for all transition events which results in exponential
distributions for the time periods during which an individual remains
in the infected ($\tau_I$) or recovered ($\tau_R$) states. 
However, as the exponential distribution is peaked at the lower end of
its support, this results in infected (recovered) individuals being most likely
to recover (lose immunity) immediately after becoming infected
(recovered), regardless of the mean
infection (recovery) period for a disease. On the other hand, the long
exponential tail of the distribution may result in some individuals
remaining in the infected (recovered) period for durations extremely
long relative to the mean period. For real diseases, on the other
hand, individuals will likely remain infected (recovered) for about
the same period - which implies that individual values of $\tau_I$ and
$\tau_R$ will be relatively tightly clustered about the corresponding
mean values~\cite{Newman2010}. 

In order to obtain distributions of $\tau_I$ and $\tau_R$ such
that the individuals are relatively more homogeneous in terms of the
infection and recovery periods (corresponding to reality) we have used
transition rates from $I$ to $R$ ($\gamma$) and from $R$ to $S$
($\mu$) that are
time-varying. In particular, the rates are functions of the time
duration for which an individual is in the infected or the recovered
state (respectively), viz.,
\begin{equation}
\gamma_i = \exp{[\beta(\delta_{i}^{I} - \tau_I)]},~~{\rm and},
\end{equation}   
\begin{equation}
\mu_i = \exp{[\beta(\delta_{i}^{R} - \tau_R)]}, 
\end{equation}   
where $\delta_{i}^{I}$ ($\delta_{i}^{R}$) is the time elapsed from the
instant an individual $i$ enters the infected (recovered) state and
$\tau_I,\tau_R$ are the average infection and recovery periods
respectively.
The tunable parameter $\beta$ determines how narrowly the resulting
distribution of periods will be spread around the corresponding mean
value. Fig.~\ref{timedistrn} compares the distributions of $\tau_I$
(top) and $\tau_R$ (bottom) obtained with a constant transition rate
(`null'), as
well as two time-varying rate processes with different values of $\beta$.
As can easily be seen, the distribution of
the individual time periods become less dispersed with increasing
$\beta$, so that events
corresponding to very short or very long infection or
recovery periods (compared to the mean values) become extremely
unlikely. For all the stochastic simulation results shown here, we
have chosen $\beta = 1$. Small variations in the value of $\beta$
about this value do not change the results qualitatively. Furthermore,
deterministic model simulations in which all individuals have the same
$\tau_I$ and $\tau_R$ (corresponding to $\beta \rightarrow \infty$)
also yield qualitatively similar results,
implying that the model behavior is not sensitively dependent on the
precise value of $\beta$ chosen as long as the distribution is
relatively peaked narrowly about the mean.

{\em Description of the simulation videos.}
\begin{figure}[tbp]
\includegraphics[width=0.98\linewidth,clip]{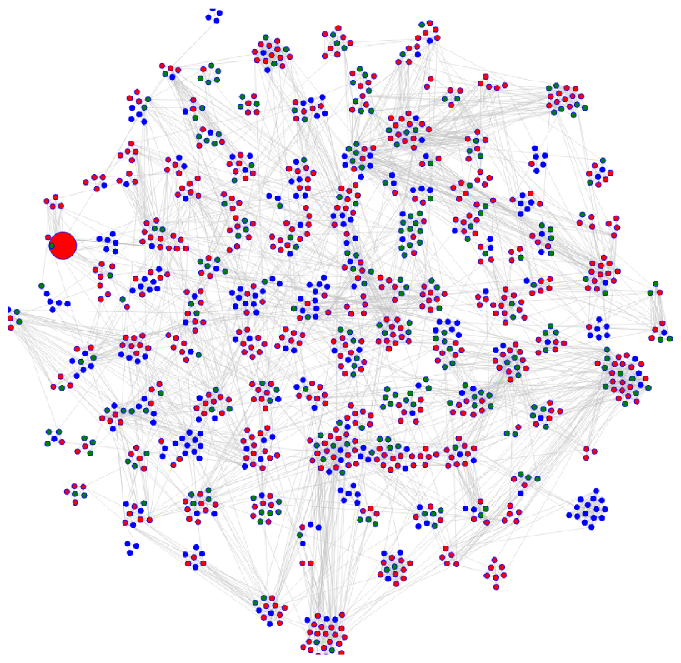}
\includegraphics[width=0.98\linewidth,clip]{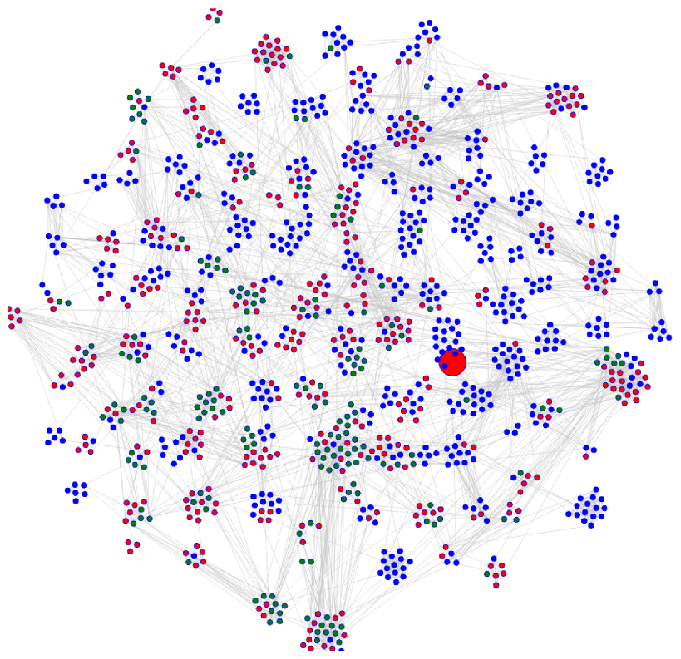}
\caption{Snapshots from the simulation videos for the scenarios in
which
(top, Video 1) the disease dies out rapidly after the initial outbreak and
(bottom, Video 2) the epidemic continues for as long as the simulation is
continued. The large red circle indicates the individual that has
become infected at the instant the snapshot is taken. Node colors
represent different states, viz., blue: susceptible, red: infected
and green: recovered.}
\end{figure}
The two videos show simulated epidemics spreading in
the social contact network of individuals in one of the 
villages in our database (Id. no. 55). 
The largest connected component of this network comprises 1151 nodes.
The videos show two contrasting scenarios that one can observe for
exactly the same choice of parameters. Video 1 corresponds to the
situation when
the infection dies out very early after initiation. Video 2 shows the
scenario where the infection persists in the network upto the entire
duration of the simulation.
The three different states the nodes can be in are represented using
three different colors,
blue representing susceptible state, red
infected state and green recovered state.
Each individual event (i.e., a node becoming infected, recovered or
susceptible) is shown with the node involved being highlighted.
The ``real time'' elapsed between two consecutive events can be of any
duration as this is a stochastic simulation.

\begin{figure}
\begin{center}
\includegraphics[width=0.9\linewidth]{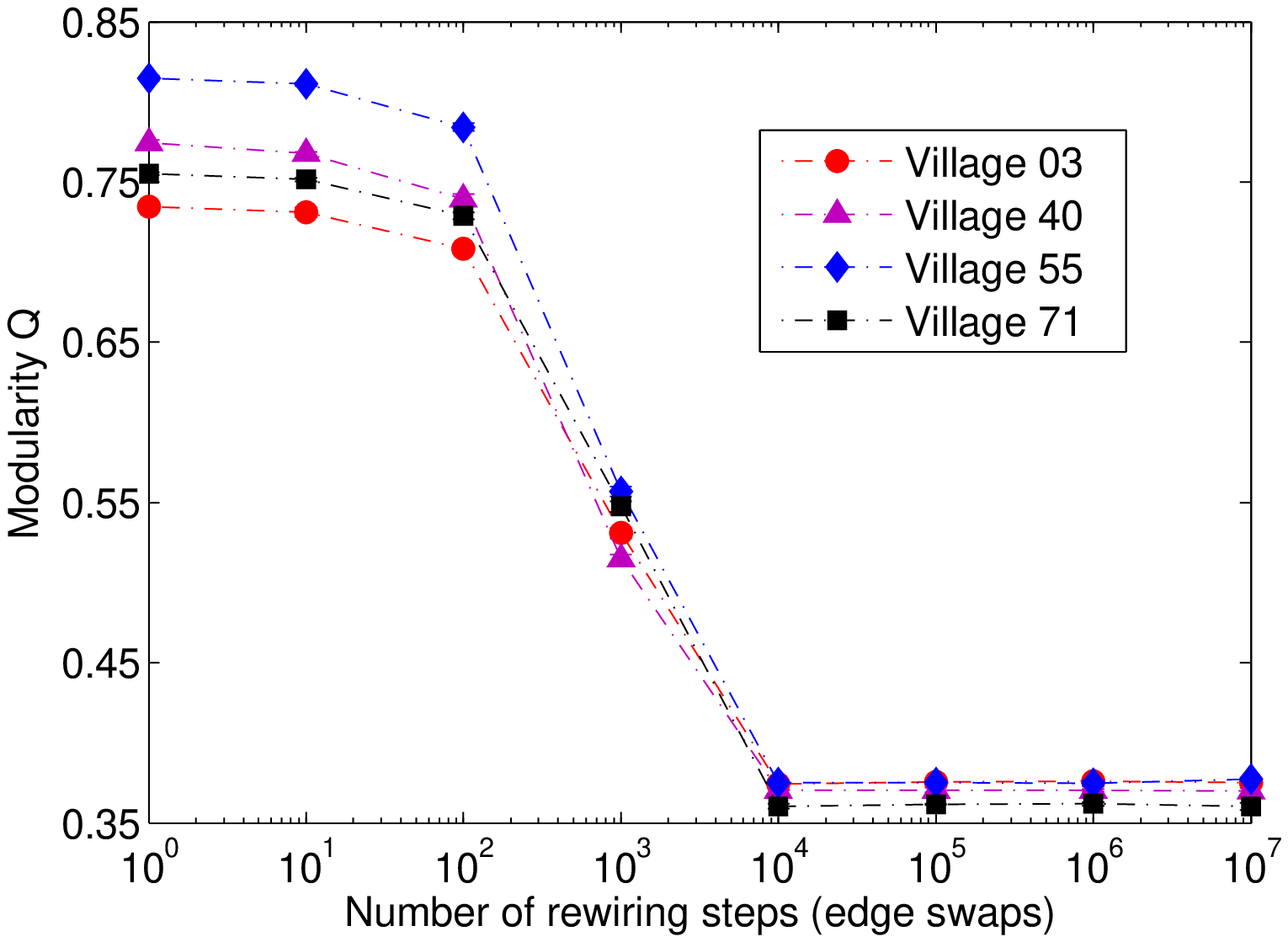}
\end{center}
\caption{The modular character of a degree-preserved randomized
network, as measured by $Q$, decreases on increasing the number
of random link exchanges starting from any one of four empirical
social networks in our database (corresponding to villages whose Id
numbers are shown in the legend). The four villages chosen for demonstration
all exhibit a relatively high level of modular organization. In all
four cases, the decrease in $Q$ saturates to an asymptotic value after
around $10^4$ link exchange steps.
}
\label{figs5}
\end{figure}

{\em Details of randomization method for empirical networks.}
In order to understand the role of modularity in empirical social
networks we need to compare them with an equivalent `randomized'
network having the same degree profile as the empirical network but
which does not have modular organization. Such a randomized network ensemble
can be constructed from the original network by a specific
link swapping procedure described as follows. We choose any two
modules of
the original network (module 1 and module 2, say) and then from each,
choose a pair of connected nodes (A,B from module 1 and C,D from
module 2, say). The link exchange corresponds to removing the two
intra-modular links and replacing them with two inter-modular links
such that the degree of each node is preserved (e.g., removing the
link between A \& B and that between C \& D, and connecting A to C
and B to D). This process when repeated many times will result in a
network whose modularity is much reduced compared to the empirical
network, while preserving the original degree profile. This is shown
in Fig.~\ref{figs5} which shows how the modular nature of a
network, as measured by $Q$~\cite{Newman06}, changes as one randomizes
empirical social networks by gradually increasing the number of
rewiring steps. We observe that the modularity of the randomized
network decreases rapidly when the number of link
exchanges is increased beyond 100, but after about $10^4$ rewirings,
$Q$ does not change appreciably. 
In our study, we have used $10^6$ link exchanges to generate
randomized network ensembles.

{\em Effect of removing only inter-modular links from the empirical
social networks.}
\begin{figure}
\begin{center}
\includegraphics[width=0.65\linewidth]{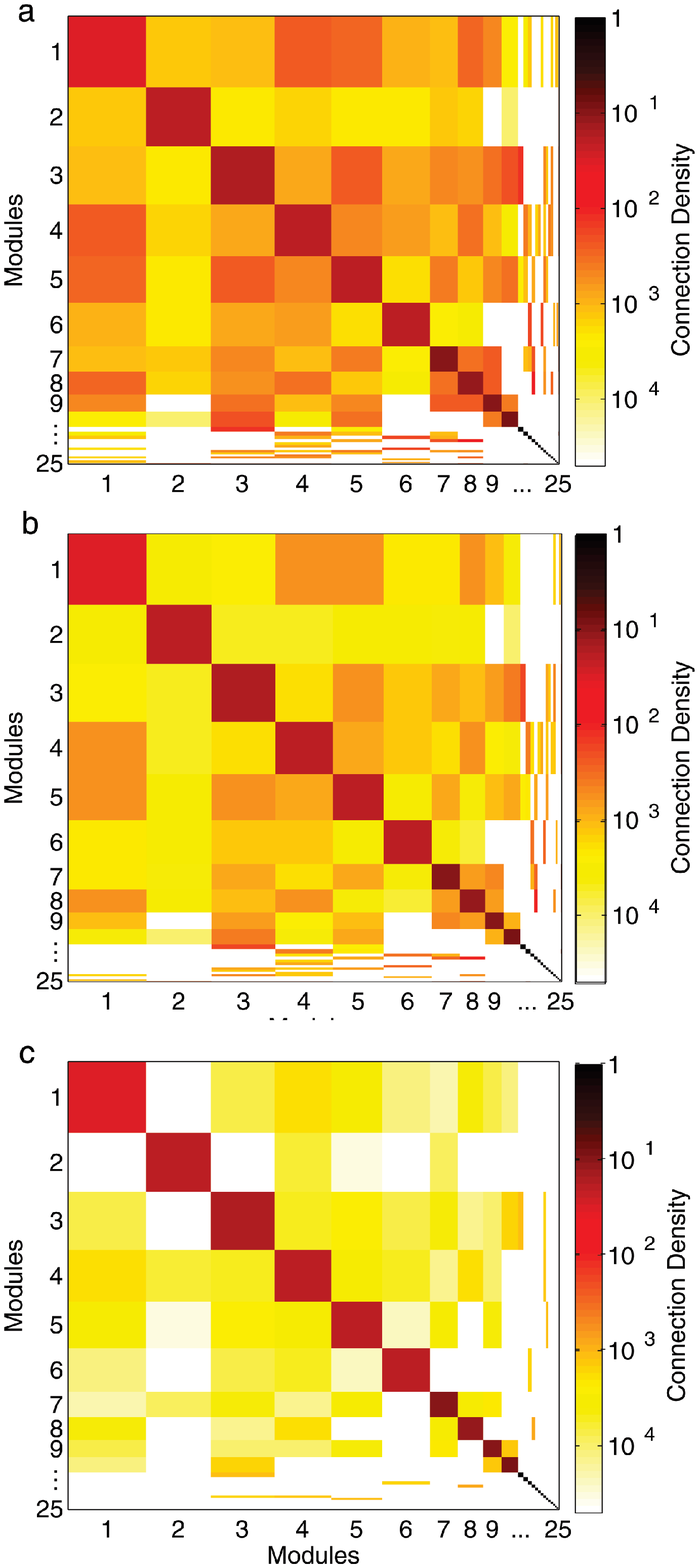}
\end{center}
\caption{Modular interconnectivity in the social network of a village
(Id. no. 52) $N=1497$
individuals.
(a) Matrix representing the average connection density between
individuals belonging to the same network community and those belonging to
different communities in the largest connected component of a village
social network of southern India (contact data obtained
from Ref.~\cite{data1}).
The communities are defined in terms of
connectivity within them being relatively denser
in comparison to the overall connection density, and have been identified
using a spectral algorithm~\cite{Newman04}. The empirical network has
been subsequently used to generate sparser contact networks by
reducing the inter-modular connection 
density, keeping the connections within each 
module unchanged. (b) and (c) show the corresponding matrices for
reduction by 50\% and 90\% respectively.
}
\label{figs2}
\end{figure}
Comparing the progress of an epidemic on an empirical social network
with community structure to that in the corresponding randomized
network that is relatively homogeneous provides an important tool to
evaluate the key role of modular organization in enhancing the
persistence of epidemic processes on networks. However, apart from
comparing the empirical networks having high modularity with
corresponding randomized networks of relatively much lower modularity,
one can also investigate networks having higher modularity that can be
constructed from
empirical social networks by selectively removing a fraction of the
links that connect nodes belonging to different modules, while
preserving all intra-modular links. An important limitation
of this procedure is that it does not preserve the degree profile of
the original network. Thus, the following results are indicative
rather than conclusive.

\begin{figure}
\begin{center}
\includegraphics[width=0.99\linewidth]{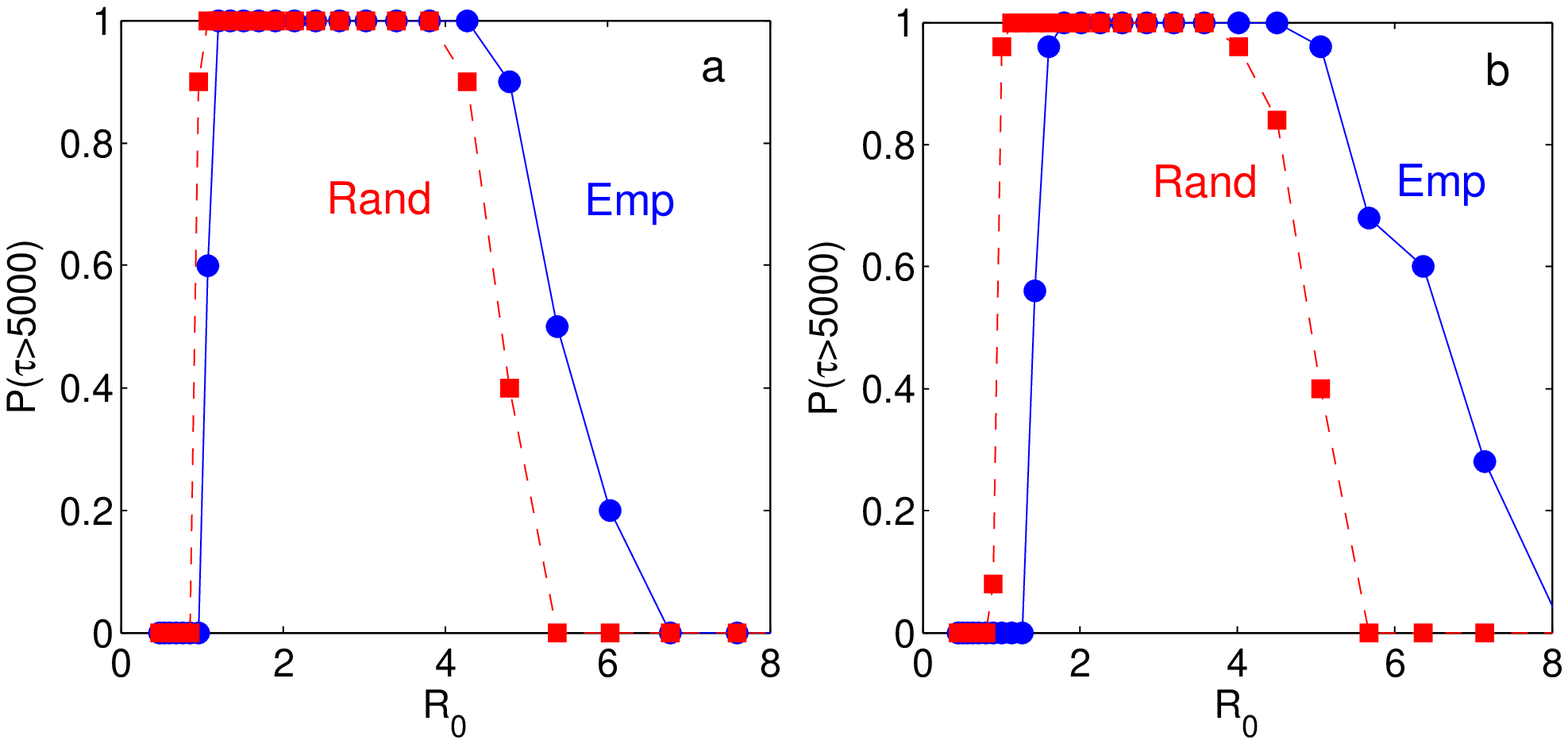}
\end{center}
\caption{The probability that persistence time of an epidemic exceeds
a specified duration (viz., 5000 time units) in modular
social networks (continuous curve) and the corresponding degree-preserved 
randomized networks (broken curves) as a function of the basic reproduction number of
the epidemic for (a) 50\% and (b) 90\% reduction of the inter-modular
connection density in the empirical contact network of a village in
southern India.
}
\label{figs3a}
\end{figure}
We show here that on increasing the modularity of the
empirical network, persistence of the epidemic can be observed for
higher values of $R_0$ compared to the original network. 
Fig.~\ref{figs2}~(a) shows the empirical network matrix,
representing the average connection density between individuals
belonging to the same network community and those belonging to
different communities. This empirical network was used to generate
higher degree of modular organization of contact networks by reducing
the inter-modular connection density by $50\%$ (b) and $90\%$ (c), keeping
the connections within each module unchanged.

Fig.~\ref{figs3a}~(a-b) shows the probability that persistence time of an
epidemic exceeds a specified duration (viz., 5000 time units) in the
reduced inter-modular connection density by $50\%$ (a) and $90\%$ (b)
empirical social network (blue continuous curve) and the
corresponding degree-preserved randomized network (red broken curve)
with the  function of $R_0$. This result demonstrates that increased
modularity in the network makes an epidemic persist for contagion
having higher level of infectiousness (i.e., higher value of $R_0$).

{\em Explaining persistence of epidemics in homogeneous networks.}
\begin{figure}
\begin{center}
\includegraphics[width=0.99\linewidth]{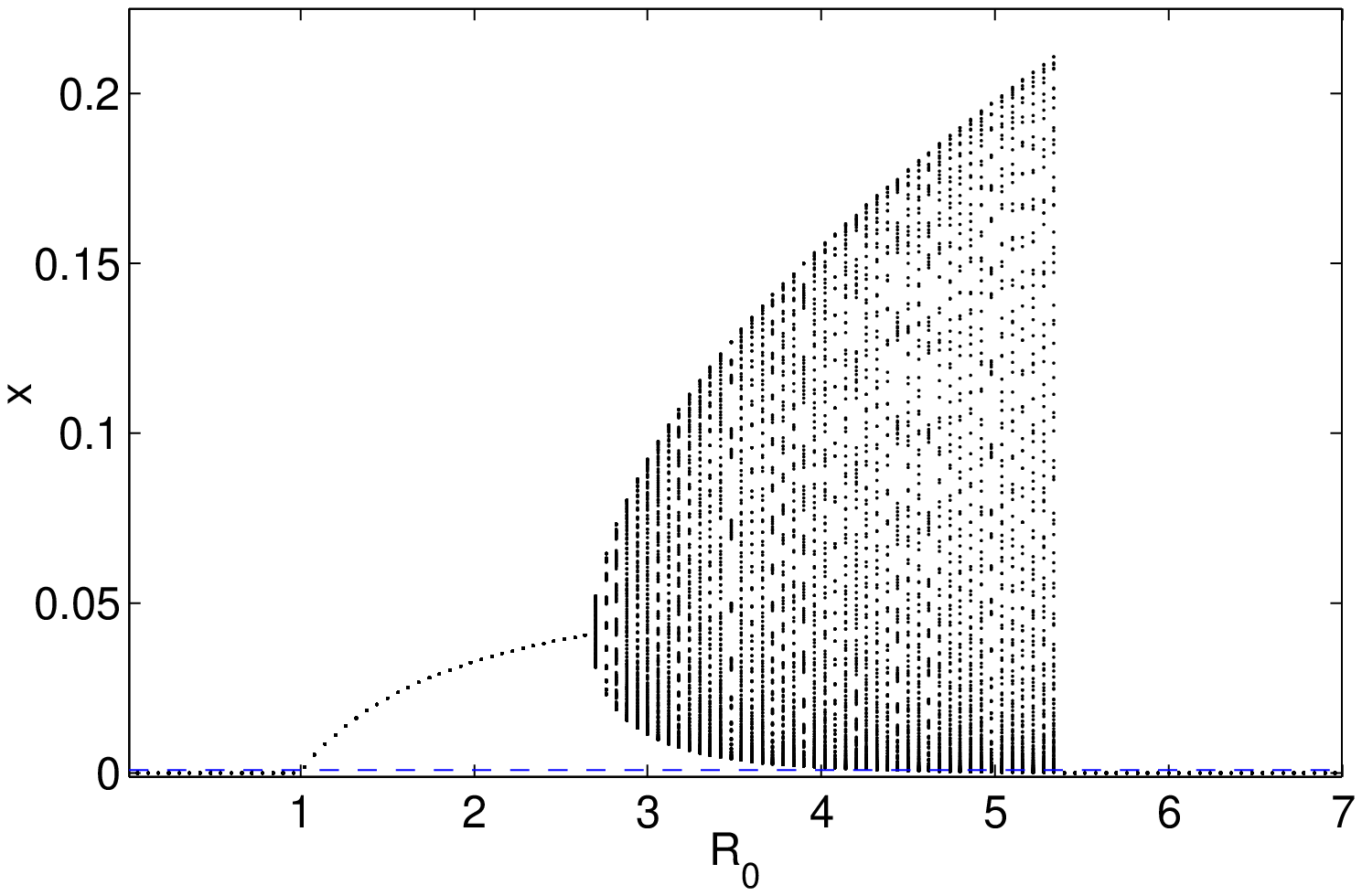}
\end{center}
\caption{Bifurcation diagram of the fraction of newly infected cases
$x_n$ at each time step $n$ in a population of $N$ individuals 
under the homogeneous mean-field
approximation such that the time-evolution of $x$ is described by
$x (n+1) = [1-(1-\beta)^{k \sum_{j=0}^{\tau_{I}-1}
x(n-j)}][1-\sum_{j=0}^{\tau_I+\tau_R-1} x(n-j)]$.
The total infected fraction in the population at time $n$ is
$f_i (n) = \sum_{j=0}^{\tau_I-1} x(n-j)$; if at any time $f_i$ falls
below the threshold $1/N$ (i.e., not a single individual in the
population is infected), indicated by the broken line, the disease
becomes extinct. This is indeed seen to occur for $R_0 \simeq 5.4$.
For consistency with results shown in Fig.~\ref{fig2}, we have chosen
$N = 1024$, $k=12$, $\tau_I =5$ and $\tau_R =10$.
In the initial state $1\%$ of the population is infected after which
the system is evolved for $10^4$ time steps for 
each value of $R_0$. In the diagram the
final $10^3$ states in the time-evolution of $x$ are shown.
}
\label{figs4}
\end{figure}
The persistence behavior of epidemics for an intermediate range of
$R_0$ can be understood in the limit of a homogeneous network, as one
can invoke a mean-field approximation that allows us to give an effective
low-dimensional dynamical description of the system.
We note that a disease having $R_0 < 1$ will not be able to initiate
an epidemic
regardless of the network structure, as the number of secondary
infections arising through contact are actually less than the number
of initially infected individuals. Thus, as the infected fraction of
the population $f_i (=I/N)$ rapidly decays to zero, the time for
which the disease persists in the population is typically short.  For
$R_0 \geq 1$, relatively homogeneous networks (i.e., having higher
values of $r$) show a rapid rise in the disease incidence
characteristic of an epidemic as the network spreads the infection to
a much larger fraction than that which had been infected initially.
The infected fraction time-series then settles down to an irregular
series of oscillations with the disease persisting in the population
for the duration of simulation, provided $R_0$ is not too high.
However, if $R_0$ is increased, we eventually observe a
very different long-term behavior where the entire population becomes
infected in a short space of time, followed by recovery and extinction
of the disease. This can be explained in terms of a delay difference
equation describing the time-evolution of infections under a
mean-field approximation (where we can assume $k_{inf} = k f_i$).
For such a case, the fraction of individuals that are infected at a
particular time-step $n+1$ is $x_{n+1} = [1-(1-\beta)^{k
\sum_{j=0}^{\tau_{I}-1} x(n-j)}][1-\sum_{j=0}^{\tau_I+\tau_R-1}
x(n-j)]$.
As can be seen from Fig.~\ref{figs4}, the bifurcation diagram of the
map shows that for an intermediate range of $R_0$ the epidemic will be
persistent.

{\em Construction procedure for model networks.} 
The modular network model used in this paper follows 
from the definition of modularity and
consists of $N$ nodes arranged into $M$ modules of different sizes.
The size distribution of the modules has a Gaussian nature, whose
dispersion can be tuned around an average
value $n = N/M$, by changing the standard deviation $\sigma$.
The connection probability between nodes in
a module is $\rho_i$, and that between different modules is $\rho_o$.
The parameter defining the model is the ratio of inter- to
intra-modular connectivity, $\rho_o/\rho_i = r \in [0, 1]$. For
$r \rightarrow 0$, the
network gets fragmented into isolated clusters, while as
$r \rightarrow 1$, the network approaches a homogeneous random network.
If $n_i$ is the size of module $i$ and $k$ is the average degree, 
the intra- and inter-modular connection densities can be expressed in
terms of the different network parameters as
$\rho_i = N k / [\Sigma_i n_i (n_i - 1) + \Sigma_i r n_i (N - n_i)]$
and $\rho_o = r \rho_i$. The adjacency matrix for a contact network is
constructed by randomly linking nodes within a module and between
modules according to the above connection probabilities, respectively.
Details will be provided in a forthcoming publication~\cite{Menon15a}.

{\em Persistence of epidemics in modular network model.}
\begin{figure}
\begin{center}
\includegraphics[width=0.99\linewidth,clip]{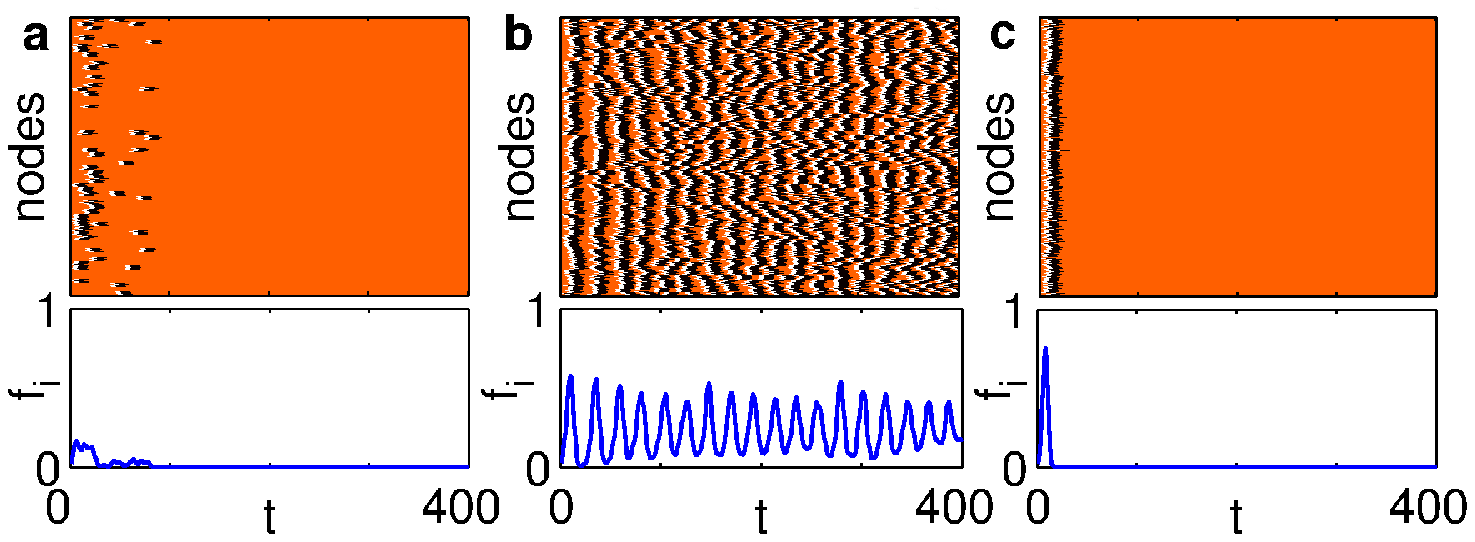}
\end{center}
\caption{Optimal modular organization of the contact network makes
an infectious disease persistent. 
Space-time diagrams (top row) and the
time-series (bottom row) of the infected fraction $f_i$ of population
shown for contact networks having (a) $r = 2
\times 10^{-4}$, (b) $r = 2 \times 10^{-3}$ and (c) $r = 2
\times 10^{-2}$, for a disease with $R_0 = 6$. While both for the case
of almost isolated modules (a) and the relatively homogeneous network
(c), the epidemic
becomes extinct within 100 time units, for an intermediate value of
modularity
organization (b) the epidemic persists for as long as the simulation
is continued.
For all simulation results shown here we have used a network of
$N=1024$ nodes having $M=64$ modules of size $n=16$ each. The nodes
have average degree $k=12$, with $\tau_I =5$ and $\tau_R =10$ time
units.}
\label{figs6}
\end{figure}
\begin{figure}
\begin{center}
\includegraphics[width=0.99\linewidth,clip]{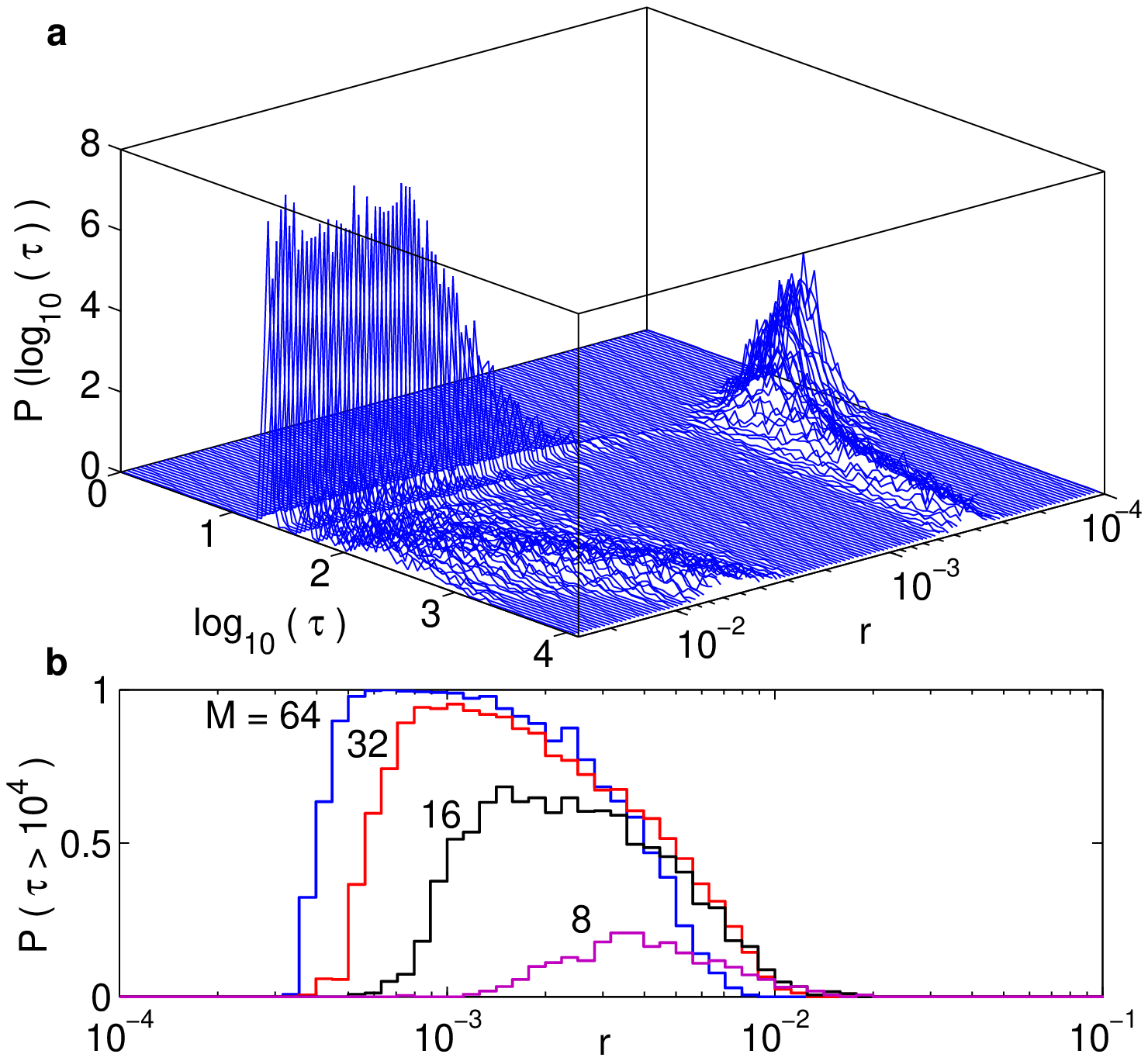}
\end{center}
\caption{
(a) The distribution
of the time $\tau$ (with logarithmic binning) for which an epidemic
with $R_0 = 6$ 
persists in the network shows a bimodal nature for higher values of
the modularity parameter $r$, with the upper branch diverging as the
network becomes more modular. For lower values of $r$, the
distribution is unimodal and the average value of $\tau$ decreases
rapidly as the modules become effectively isolated. Results shown for
$N=1024$ with $M=64$ modules, each with $n=16$ nodes having an average
degree $k=12$.
(b) The probability that the epidemic persists for more than $10^4$
time units is shown as function of the modularity parameter $r$ for
networks having different number of modules $M$ ($N=1024$ with each
node having average degree $k=12$.)
}
\label{figs3}
\end{figure}
As already described in the main text, an epidemic that becomes
extinct within a short time interval in a homogeneous network ($r \sim
1$), can
become recurrent when the network is optimally modular,
persisting in the system
for extremely long times. However, as one approaches
the limit of the modules becoming almost isolated ($r \sim 0$), the
infection is unable to transfer from one module to another and the
epidemic again becomes extinct rapidly.
This is shown explicitly in Fig.~\ref{figs6}. 

Fig.~\ref{figs3}~(a) shows the variation of the probability distribution
of persistence time $\tau$ with network modularity.
The distribution of $\tau$ over the range of $r$ for which the
epidemic persists indefinitely shows a bimodal nature. The lower
branch corresponds to the time-scale over which the disease lasts in a
single module, and as can be seen the peak of the distribution remains
almost unchanged as $r$ is varied. For low $r$, it represents the
realizations in
which the disease fails to break out of the module(s) in which
the initial outbreak occurs. For high $r$, the contagion spreads
rapidly through the system, lasting almost as long in the entire
system as it does in any given module. The upper branch, on the other
hand, corresponds to the realizations where the epidemic propagation
between modules is slowed down by the modular organization. Note that
the peak of this branch diverges in the optimal range of $r$ where the
epidemic becomes persistent in the network. Moreover, the bulk of the
distribution moves from the lower to the upper branch in this
parameter range. 
Fig.~\ref{figs3}~(b) shows that while the optimal modularity range is not
overly dependent on the number of modules $M$ in the network, the
persistence effect does
become more prominent as one increases $M$, while keeping the total size of the
network ($N$) invariant. We also observe that the value of $r$ at
which the peak of the persistence behavior occurs reduces
with increasing $M$.

\begingroup
\squeezetable
\begin{table*}
\caption{Summary details of the
contact network properties for the villages considered in this study.}
\begin{ruledtabular}
   \begin{tabular}{c c c c c c c c c c}

SlNo&Nodes&LCC&$<k>$&Modules&Q&$\mu_{sz}$&$\sigma_{sz}$&LMS&r\\ \hline
1&843&825&8.2&24&0.6782&34.375&19.6783&72&0.02\\ 
2&877&810&7.2123&23&0.7396&35.2174&22.446&91&0.0149\\ 
3&1380&1318&7.9841&25&0.7312&52.72&38.6539&189&0.0161\\ 
4&1025&957&7.2403&25&0.7315&38.28&25.6117&99&0.0155\\ 
5&650&641&7.869&25&0.7477&25.64&20.268&70&0.0154\\ 
6&451&434&6.9954&24&0.7036&18.0833&15.4809&63&0.0193\\ 
7&732&719&9.0668&23&0.7078&31.2609&25.1555&83&0.0195\\ 
8&444&440&8.2818&22&0.7051&20.0&20.3693&75&0.0222\\ 
9&928&914&8.4497&22&0.703&41.5455&33.3517&141&0.0212\\ 
10&354&346&8.8353&19&0.6614&18.2105&15.2784&48&0.0294\\ 
11&605&589&7.8727&23&0.6986&25.6087&23.814&92&0.023\\ 
12&794&760&7.7132&23&0.7225&33.0435&20.0531&74&0.0167\\ 
13&-&-&-&-&-&-&-&-&-\\ 
14&675&645&8.1054&23&0.7453&28.0435&21.1115&91&0.0155\\ 
15&853&852&8.9648&23&0.6954&37.0435&24.5985&90&0.0204\\ 
16&712&693&9.3059&22&0.7152&31.5&24.7694&99&0.0201\\ 
17&879&850&8.6494&25&0.7284&34.0&28.1539&123&0.0177\\ 
18&1146&1140&9.157&24&0.7522&47.5&34.87&127&0.0149\\ 
19&1134&1118&9.2844&24&0.7484&46.5833&31.0979&131&0.014\\ 
20&714&633&8.6477&23&0.744&27.5217&18.5117&66&0.0139\\ 
21&1046&1011&8.6311&25&0.7347&40.44&32.0063&116&0.0164\\ 
22&-&-&-&-&-&-&-&-&-\\ 
23&1252&1186&8.6636&24&0.7572&49.4167&37.5732&125&0.0144\\ 
24&835&820&9.3098&25&0.7222&32.8&22.735&79&0.0165\\ 
25&1313&1286&9.4619&25&0.7751&51.44&32.2404&124&0.0116\\ 
26&674&666&9.0405&21&0.7576&31.7143&28.3198&127&0.0152\\ 
27&708&682&7.4091&25&0.7408&27.28&18.3794&67&0.0144\\ 
28&1612&1570&9.5822&25&0.7812&62.8&39.8628&163&0.0112\\ 
29&1337&1270&7.8276&25&0.7651&50.8&26.9741&115&0.0113\\ 
30&689&675&8.8607&23&0.7489&29.3478&25.1441&91&0.0165\\ 
31&851&819&9.0317&25&0.7893&32.76&22.0023&97&0.0106\\ 
32&1181&1136&9.6514&25&0.7296&45.44&30.0281&102&0.0157\\ 
33&843&824&7.7415&25&0.7279&32.96&29.0799&101&0.0174\\ 
34&692&628&7.1385&24&0.7702&26.1667&26.1465&115&0.0128\\ 
35&806&756&7.2302&25&0.758&30.24&27.388&106&0.0133\\ 
36&1214&1168&8.8733&24&0.7147&48.6667&53.5698&197&0.0214\\ 
37&500&482&7.7759&16&0.6593&30.125&24.2922&91&0.0324\\ 
38&736&726&8.1267&24&0.7723&30.25&21.9037&82&0.0124\\ 
39&1339&1294&9.1376&25&0.7666&51.76&31.9466&134&0.012\\ 
40&1097&1064&8.0442&25&0.7713&42.56&42.0562&168&0.0135\\ 
41&724&703&7.8862&20&0.7163&35.15&26.3615&76&0.0203\\ 
42&853&805&8.0807&25&0.7587&32.2&32.3394&117&0.0153\\ 
43&875&861&8.295&24&0.7331&35.875&30.5195&121&0.016\\ 
44&978&965&8.8518&25&0.7212&38.6&40.4366&136&0.0202\\ 
45&1073&1044&8.2356&25&0.7782&41.76&29.4405&113&0.0116\\ 
46&1257&1216&7.8544&24&0.7683&50.6667&34.6671&155&0.0125\\ 
47&680&660&8.5848&24&0.719&27.5&20.4185&70&0.0174\\ 
48&808&794&8.9232&24&0.6998&33.0833&31.5686&116&0.0225\\ 
49&766&689&8.7083&22&0.6439&31.3182&35.6987&140&0.0362\\ 
50&999&937&8.9883&25&0.6994&37.48&43.5175&145&0.0235\\ 
51&1061&1015&10.6591&21&0.6734&48.3333&55.2943&187&0.0308\\ 
52&1525&1497&10.4369&25&0.7339&59.88&69.595&276&0.0192\\ 
53&642&630&9.0683&23&0.6573&27.3913&32.5746&110&0.0322\\ 
54&467&458&10.1528&20&0.6636&22.9&28.0141&111&0.0326\\ 
55&1180&1151&7.9644&24&0.8192&47.9583&33.5478&127&0.0087\\ 
56&573&553&7.8807&23&0.7223&24.0435&20.6955&76&0.0189\\ 
57&948&919&8.2459&25&0.7222&36.76&29.2278&130&0.0174\\ 
58&914&905&8.8541&25&0.744&36.2&27.0289&110&0.0153\\ 
59&1599&1552&8.5393&24&0.7936&64.6667&46.4091&182&0.0108\\ 
60&1775&1729&8.7953&25&0.789&69.16&42.0692&215&0.0105\\ 
61&591&572&9.3776&20&0.7299&28.6&25.4213&104&0.0191\\ 
62&994&980&9.1816&24&0.7676&40.8333&30.4872&131&0.013\\ 
63&786&774&8.0284&23&0.7833&33.6522&23.392&91&0.0119\\ 
64&1286&1265&8.7763&25&0.7926&50.6&32.3988&115&0.0103\\ 
65&1331&1301&9.3766&24&0.7101&54.2083&42.7444&149&0.0194\\ 
66&814&790&7.8848&23&0.7353&34.3478&29.078&102&0.0173\\ 
67&893&885&8.5401&22&0.7311&40.2273&36.4778&128&0.0192\\ 
68&663&655&8.1389&22&0.7014&29.7727&26.6218&90&0.0225\\ 
69&875&866&10.4688&23&0.6739&37.6522&42.5457&148&0.0281\\ 
70&899&891&9.3547&20&0.6634&44.55&37.4639&129&0.028\\ 
71&1387&1345&8.3836&24&0.7556&56.0417&44.8046&151&0.0144\\ 
72&999&977&8.6192&25&0.6973&39.08&33.5511&147&0.0217\\ 
73&870&858&9.4767&22&0.7207&39.0&31.7905&101&0.0199\\ 
74&743&724&8.2735&23&0.7707&31.4783&22.9307&89&0.0131\\ 
75&831&815&10.0454&24&0.7199&33.9583&43.6305&170&0.0236\\ 
76&1154&1126&8.3064&25&0.7878&45.04&34.5861&128&0.0114\\ 
77&707&671&8.2355&22&0.7551&30.5&24.5352&82&0.015\\
\end{tabular}
\end{ruledtabular}
\label{table1}
\end{table*}
\endgroup

\end{document}